\documentstyle{article}
\begin{document}
\title{\bf Quantum Mechanics for Totally\\
 Constrained Dynamical Systems\\
and Evolving Hilbert Spaces}
\author{Ricardo Dold\'an, Rodolfo Gambini and Pablo
Mora\\
Instituto de F\'{\i}sica, Facultad de Ciencias \\
Trist\'an Narvaja 1674, Montevideo, Uruguay}
\date{\today}
\maketitle
\begin{abstract}
We analyze the quantization of dynamical systems that do not
involve any background notion of space and time. We give a set of
conditions for the introduction of an intrinsic time in quantum
mechanics. We show that these conditions are a generalization of
the usual procedure of deparametrization of relational theories
with hamiltonian constraint that allow to include systems with an
evolving Hilbert Space. We apply our quantization procedure to the
parametrized free particle and to some  explicit examples of
dynamical system with an evolving Hilbert space. Finally, we
conclude with some considerations concerning the quantum gravity
case.

\end{abstract}

\section{Introduction}

The concept of time enters in the basic formalism of quantum
mechanics in two ways: to mark the evolution of the system and to
order a sequence of measurements. In terms of Von Neumann
\cite{von} axiomatic formulation time enters as evolution labeling
parameter in axiom IV through the evolution equation (Schroedinger
equation) and implicitly in axiom II through the possible
dependence of the operators corresponding to observables on time.
On the other hand time appears as sequence ordering label in axiom
V, through the fact that the outcome of a measurement depends on
previous measurements. Furthermore this time parameter is assumed
to be given in advance. The picture that we get is a unit vector in
a Hilbert space (which depends on the system and is given once and
forever, following axiom I) with a smooth time evolution generated
by the hamiltonian operator via the Schroedinger equation with
discontinous leaps corresponding to measurements.\\

The Dirac \cite{Dirac} quantization procedure for constrained
systems doesn't introduce major changes in this picture,
considering that it assumes the existence of a non vanishing
hamiltonian in addition to the set of constraints, and a standard
Schroedinger equation with that hamiltonian. There exist however a
wide class of models for which, at the end of the application of
the usual thumb rules of quantization, one is left only with a set
of constraint equations (in addition, of course, of the commutation
algebra of the fundamental dynamical variables), without neither a
non zero hamiltonian nor a natural choice of a time parameter. This
situation is characteristic for instance of reparametrization
invariant systems (see \cite{sunder}), sometimes called generally
covariant systems. The example of greatest physical interest of
this kind of theories is undoubtely General Relativity, where the
problem is known as "the issue of time" \cite{Kuc1,Isham1}.\\

Our aim in this paper is twofold. Firstly we intend to propose the
necessary changes in the standard formalism of quantum mechanics in
order to deal with the above mentioned kind of systems, which we
will call "totally constrained systems". This involve to put
forward a prescription to slice the representation space in which
we will realize the commutation algebra of the dynamical variables
in equal "time" spaces as well as define this "time". Then we will
explore the logical possibility of the slices being non isomorphic.
This could be considered, from the point of view of standard
quantum mechanics, as a change in time or evolution of the Hilbert
space describing the system.\\

The motivation to consider the possibility of "evolving Hilbert
spaces" comes from the conceptual point of view from the
suggestion, due to Unruh \cite{Unruh}, that quantum gravity should
have this property .The main point of his argument is the
following: Does the Big Bang theory  for the origin of the Universe
means that because there was less space early and there were also
fewer physical attributes that the Universe had? His answer is yes
and it is based in the fact that there should be some limit at the
Planck length to the number of different values that any field
could take. If this observation is true one should describe the
universe with a finite dimensional Hilbert space and a set of
operators which both change in time\footnote{A similar proposal was
made by Jacobson \cite{Jac}}. This proposal seem to be very
appealing both from a physical and philosophical point of view. In
fact in a description of the universe in terms of fixed Hilbert
space, the set of possible behaviors of the universe is fixed at
all times from the very beginning. That means that the state that
describes the present behavior of the universe with its enormous
complexity was a vector of the Hilbert space since the Big Bang. In
practice the observed evolution from the simple to the complex is
nothing but the evolution between different possible behaviors. In
quantum mechanics a system is identified with its Hilbert space,
the set of all its  possible behaviors (states). Hence, in this
picture, the universe is given once and for all.

Furthermore, if the Hilbert space is fixed, the initial conditions
of the universe are not determined by its dynamical laws and the
actual initial conditions remains completely unexplained. Hartle
\cite{Hartle} has stressed the reasons for the search of a theory
for the initial conditions of the universe. Initial conditions are
crucial to explain the large scale homogeneity and isotropy of the
universe,its approximate spatial flatness, the spectrum of density
fluctuations, the homogeneity of the arrow of time and the
existence of a classical space time.

 Of course, the usual quantum mechanical systems like particles are
described by a fixed set of attributes as position and momentum and
a fixed Hilbert space, but for these systems we have no reason to
expect different behaviors for different  times,  and  in
principle  any conceivable initial state may be prepared by
measuring a suitable CSCO.

In this work, as we said before, we are interested in the
identification of an intrinsic time in totally constrained
dynamical systems. We shall give a set of conditions for the
definition of a physical time that generalize the usual
deparametrization procedure.We shall see that the introduction of a
physical time in these systems naturally leads to the possibility
of evolving Hilbert spaces.

Some systems that usually require the introduction of a nonpositive
definite inner product or a decomposition between positive and
negative frecuency states may now be quantized in terms of a
positive definite inner product with an evolving Hilbert space. In
general, evolving Hilbert spaces seem to be naturally related with
systems with boundaries or involving operators that satisfy a non
canonical algebra.

As it was noticed by Unruh, evolving Hilbert spaces are naturally
related with systems with a finite number of degrees of freedom. In
fact, in the infinite dimensional case, it is always possible to
describe the system in terms of a fixed Hilbert space, but we shall
prove that,in this case, relational systems may behave as the
continuum limit of systems with a finite dimensional evolving
vector space. In particular, the transition amplitudes will remain
invariant while the system evolves into the future , but the system
will not be invariant under time reversal, and the evolution will
not be unitary. We shall call this kind of infinite dimensional
systems evolving systems.

In section 2 we introduce a description of the quantum mechanics of
totally constrained dynamical systems and show that this
description naturally generalizes the quantization procedure of
deparametrizable systems. In section 3 we are going to apply this
description to three examples. The parameterized classical free
particle, a finite dimensional constrained system with an evolving
vector space of states and an infinite dimensional evolving system
associated with the Klein Gordon model.

Finally,in section 4, we shall conclude with some final remarks
concerning the application of this procedure to the quantum gravity
case.

\section{Time and quantization of totally constrained dynamical
systems}

We will assume that we have, in general as the outcome of a standard
hamiltonian formulation of the theory under study, a set of
constraint equations
\begin{eqnarray}
\phi _i(q_a)=0\,\,\,\hbox{\rm with}\, i=1,....,n+1
\end{eqnarray}
written in terms of the dynamical variables $q_a$, with
$a=1$,...,$f$ of the theory as well as the commutation (or
anticommutation) algebra of this variables

\begin{eqnarray}
[q_a,q_b]_{\pm}={\alpha}^c_{ab} q_c+{\beta}_{ab}
\end{eqnarray}
Notice that if
\begin{eqnarray}
{\alpha}^c_{ab}=0\\
{\beta}_{ab}=i
\end{eqnarray}
then $q_a$ and $q_b$ are canonically conjugated variables.
From the algebra of the $q$'s it follows the algebra of the
constraints

\begin{eqnarray}
[\phi _i,\phi _j]_{\pm}=f^k_{ij}(q_a)\phi _k
\end{eqnarray}

which we assume closed (or the constraints to be first
class) , but
allowing the "structure constants" to depend on the $q$'s. We will
assume furthermore that one of the constraints is singled out (as
it is the case in concrete examples) from a physical point of view
or considerations from the classical theory as containing
implicitly the information about the time evolution of the system,
and we will call it {\it hamiltonian constraint} ${\cal H}$. We
will call the remaining $n$ constraints {\it kinematical
constraints} ${\cal D}_i$.

In order to have a sensible quantum theory out of the previous
information we need to fulfill several steps.

The first ones, more or less straigthforward, are:

(1) To find a vector space ${\cal E}_R$, which we will call {\it
representation space}, in which realize the algebra of the $q$'s

(2) To construct the subspace ${\cal E}_K$, which we will call {\it
kinematical space} given by the solutions $\mid\psi _K\rangle$  of
the whole set of kinematical constraints

\begin{eqnarray}
{\cal D}_i\mid\psi _K\rangle =0
\end{eqnarray}

(3) To construct the subspace ${\cal E}_F$, which we will call {\it
physical space} given by the solutions $\mid\psi _F\rangle$  of the
whole set of constraints \begin{eqnarray} {\cal D}_i\mid\psi
_F\rangle =0\\ {\cal H}\mid\psi _F\rangle =0 \end{eqnarray}

At this point we will consider a function $T(q_a)$ of the dynamical
variables and we will discuss which conditions should be satisfyed
by this function in order to be considered a time variable for the
system.

{\bf (I)} The first condition that we will require is
\begin{equation}
[T, {\cal D}_i] \,=\,0\; \hbox{and}\;[T, {\cal H}] \;\ne\,0
\end{equation}
\vskip 0.5 cm
The vanishing of the commutator between $T$ and ${\cal D}_i$ imply
that $T$ is a well defined operator in ${\cal E}_K$
($T\mid\psi _K\rangle\in{\cal E}_{K}$), while the nonvanishing
of the commutator
between the hamiltonian constraint and $T$, together with conditions
III and IV insure that the
hamiltonian
restrict the evolution in $t$ of the wavefunctions. This condition is
not independent of the other conditions . In fact, if the hamiltonian
commutes with T condition IV obviously fails.

{\bf (II)} We will require that the eigenvectors of $T$ span the
kinematical space ${\cal E}_{K}$. In other words, there exist a
basis $\mid x, t\rangle$ of ${\cal E}_{K}$ such that

\begin{equation}
T\mid x, t\rangle =t\mid x, t\rangle
\end{equation}
where the $x$ correspond to additional labels necessary to
characterize the vector unambiguously.

\vskip 0.5 cm

We shall consider the vector space ${\cal E}_{Kt}$ defined  as the
subspace of ${\cal E}_{K}$ spanned by $\mid x,t\rangle$ for a given
$t$. One can define an analogous vector space ${\cal E}_{Ft}$ with
the projection of the vectors of the physical space

\begin{equation}
\mid\psi _F\rangle =\sum _{xt}\psi _{Ft}(x,t)\mid x,t\rangle
\end{equation}
for a given $t$
\begin{equation}
\mid\psi _{F},t\rangle =\sum _{x}\psi _{Ft}(x,t)\mid x,t\rangle
\end{equation}

${\cal E}_{Ft}$ may be considered as the projection of ${\cal
E}_{F}$ on ${\cal E}_{Kt}$. We will introduce evolving systems by
considering the logical possibility that the components of a given
$\psi _F(x,t)$ of vectors in the physical space $\mid\psi
_F\rangle$ might vanish identically for $t< t_0$. Thus, one may
classify these vectors by the value $t_0$. We shall say that
$\mid\psi^{t_0}_F\rangle$ is from level $t_0$ if and only if

$$\psi ^{t_0}_F(x,t)\equiv 0\,\,\, \forall \, t<t_0$$ 

We will consider now the operators $O_i(q_a)$ in ${\cal E}_K$  that
commute with all the
constraints (including the hamiltonian constraint) 
Following Kuchar we shall call these
operators  perennials.

{\bf (III)} The next condition that we are going to require is that
there exist a subset of ${\cal {E}}_{Kt}$, denoted ${\cal
{E^*}}_{Kt}$ with a positive definite inner product

\begin{equation}
<\psi_K\mid \phi_K>_t
\;=\;
\sum_x{\psi^*}_K(x,t)\phi_K(x,t)\mu(x,t)
\end{equation}

and that there is a set of perennials
such that:

(a) they are selfadjoint with this inner product, provided
their classical 
counterpart be real, and commute with
$T$, $[T,O_i]=0$.
\vskip 0.5 cm
It follows that this subset of operators doesn't mix vectors lying
in differet time sections, what imply that they are block diagonal
in ${\cal E}_{K}$, in other words
\begin{equation} O_i{\cal
E}_{Kt}\subset{\cal E}_{Kt} \end{equation} and \begin{equation}
O_i{\cal E}_{Ft}\subset{\cal E}_{Ft} \end{equation}

(b) their eigenvectors in ${\cal E}_{Ft}$  labeled by
$\mid\psi _{F_{\alpha_i}}^{t_\alpha},t\rangle$ satisfy

\begin{equation}
O_i\mid\psi_{F_{\alpha_i}}^{t_\alpha},t\rangle  =
\alpha_i\mid\psi_{F_{\alpha_i}}^{t_\alpha},t\rangle
\end{equation}
for any $t$ and $\alpha_i$ independent on $t$ .

The eigenvectors
corresponding to different eigenvalues are orthogonal.

(c) their restriction to ${\cal {E}}_{Ft}$ is a complete set of
commuting observables (CSCO).

We impose that the inner product
in ${\cal {E^*}}_{Ft}$ (induced by the inner
product in ${\cal {E^*}}_{Kt}$ ) is such that the eigenvectors of
the perennial operators satisfy the orthonormality condition

\begin{equation}
\langle\psi _{F_\alpha}^{t_\alpha}
\mid\psi _{F_\beta}^{t_\beta}\rangle _t=
\theta (t-t_\alpha)\theta (t-t_\beta)
\delta _{\alpha\beta} \label{ortho}
\end{equation}
where $\theta (t-t_\alpha)$ is the Heaviside function.

\vskip 0.5 cm
Notice that a basis in ${\cal {E^*}}_{Ft}$  includes all the
 vectors $\mid\psi _{F_\alpha}^{t_\alpha}\rangle$ of level less or
 equal to $t$.

We are going to be interested, in considering as physical
observables, not only  constants of the motion but  more general
operators. We shall call an operator $A$ an observable if and only
if

\begin{equation}
[A,{\cal D}_i]=[A,T]=0
\end{equation}
and therefore $A$ is block diagonal in
${\cal E}_{K}$, $A$
is self adjoint with respect to the inner product
and the eigenvectors of $A$ expand
${\cal E}_{Kt}$. Notice that while we have required that
any perennial commuting with T is selfadjoint,  here
one may have classical real dynamical variables that are
not associated with a selfadjoint operator and consequently
they are not observables.

{\bf (IV)}The last property that we shall require to our time
variable is that

\begin{equation}
{\cal E}^{\ast}_{Kt}\equiv{\cal E}^{\ast}_{Ft}
\end{equation}
This condition essentially implies that the hamiltonian
constraint determines
the evolution of the states but it does not restrict their
functional
dependence at a given $t$.\\
In general $A\mid \psi _F\rangle$ will not be an element of
${\cal E}_F$.
However the condition {\bf IV} allows to determine a vector of
the physical
space that coincides in ${\cal E}_t$ with any eigenvector of $A$.
Let
\begin{equation}
A\mid a_{\mu},t\rangle= a_{\mu}(t)\mid a_{\mu},t\rangle\;,
\end{equation}
then the restriction of the physical state
\begin{equation}
\mid\psi _F( a_{\mu},t_0)\rangle=
\sum_{\stackrel{\alpha}{t_\alpha\leq t_0}}
\langle\psi ^{t_\alpha}_{F_\alpha}
\mid a_{\mu},t_0\rangle _{t_0}
\mid\psi ^{t_\alpha}_{F_\alpha}\rangle
\end{equation}

to ${\cal E}_{Kt_0}$ is equal to $\mid a_{\mu},t_0\rangle$, and
then

\begin{equation}
\langle x,t\mid a_{\mu},t\rangle=
\sum_{\stackrel{\alpha}{t_\alpha\leq t}}
\langle x,t\mid\psi ^{t_\alpha}_{F_\alpha}
\rangle _t\langle\psi ^{t_\alpha}_{F\alpha}\mid a_{\mu},t\rangle _t
\end{equation}
Notice that from condition {\bf III} it follows that
$\mid\psi _F( a_{\mu},t_0)\rangle$
exists and is unique,
therefore we know how to compute the transition amplitudes between
the eigenvectors of two observables $A$ and $B$ at different times

\begin{equation}
\langle b_{\nu},t'\mid\mid a_{\mu},t\rangle=
\sum_{\stackrel{\alpha}{t_\alpha\leq t}}\langle b_{\nu},t'
\mid\psi ^{t_\alpha}_{F\alpha}
\rangle _{t'}\langle\psi ^{t_\alpha}_{F\alpha}
\mid a_{\mu},t\rangle _t
\end{equation}
where we have introduced the notation $\langle\mid\mid\rangle$
to distinguish the
transition amplitudes from ordinary inner products at $t$.
In other form we have
\begin{equation}
\langle b_{\nu},t'\mid\mid a_{\mu},t\rangle=
\langle\psi _F(b_{\nu},t')\mid\psi _F(a_{\mu},t)\rangle _{t'}
\end{equation}
These amplitudes contain all the basic information required
to determine
the evolution of the system. Notice that within this context
neither all the perennials are observables
nor all the observables are perennials.\\

The state $\mid \psi_F(a_\mu,t_0)\rangle$ has been prepared by the
measurement of the observable $A$ at time $t_0$. Notice that two
states prepared at times $t_0$ and $t_1$ are such that $\langle
\psi_F(a_\mu,t_0)\mid \phi_F(b_\nu,t_1)\rangle$ is time independent
for all $t \ge t_0$ and $t \ge t_1$, {\em this is the "unitarity"
condition for an evolving system}.\\

In the next section, we shall see that these conditions define
a natural extension of the deparametrizable systems.

\section{Deparametrizable Models}

In this section we want to determine the set of necessary and
suficient conditions that a totally constrained dynamical system
should obey in order to be deparametrizable. We are going to prove
that the set of conditions given in the previous section contain as
a particular case the deparametrizable systems. Consecuently our
formalism may be considered as an extension of the usual
quantization procedure for deparametrizable systems. In a
deparametrizable model there is a (non-canonical) transformation
leading from the original set of dynamical variables $q_a$, to a
new set of variables $T$, $p_T$ and $k_a$ $a=1,...,f-2$ satifying
the algebra

\begin{eqnarray}
[k_a,k_b]_{\pm}={\alpha}^c_{ab} k_c+{\beta}_{ab}\\
{[T , p_T ]}_- =i\\
{[k_a,T ]}_- ={[k_a, p_T ]}_- =0
\end{eqnarray}
such that the hamiltonian constraint takes the form
\begin{eqnarray}
{\cal H}=p_T+H(k_a,T)
\end{eqnarray}

The other kinematical constraints have the form
\begin{equation}
\phi_j(k_a,T)=0 \label{vinc}
\end{equation}
Here we shall restrict our analysis to the case where the only
constraint is the hamiltonian constraint. The generalization of the
following considerations to the case in which there is a set time
independent kinematical constraints $\phi_j(k_a)=0$ is
straightforward.

From the algebra it follows that we can write the representation
space as a tensor product of two spaces
${\cal E}_R\equiv{\cal E}_T\otimes {\cal E}_Q$ in which
one realize on one hand $T$
and $p_T$ and in the other hand the $k_a$'s. We will chose the basis
\begin{eqnarray}
\mid x,t\rangle =\mid x\rangle\mid t\rangle
\end{eqnarray}
where $x$ correspond to the set of labels required to specify the
vector in ${\cal E}_Q$. An arbitrary vector in ${\cal E}_R$ will be
\begin{eqnarray}
\mid\psi\rangle =\sum_{xt}\psi (x,t)\mid x\rangle\mid t\rangle
\end{eqnarray}
A positive definite inner product is introduced in ${\cal E}_Q$

\begin{equation}
<\psi\mid \phi>
\;=\;
\sum_x{\psi^*}(x,t)\phi(x,t)      \label{in}
\end{equation}

such that $H$ is a hermitian operator in ${\cal E}_Q$ .

The action of the operators will be
\begin{eqnarray}
T\psi (x,t) =t\psi (x,t)\\
p_T\psi (x,t) =
-i\frac{\partial}{\partial t}\psi (x,t)\\
k_a\psi (x,t) =
\sum_{x'}(k_a)_{xx'}\psi (x',t)
\end{eqnarray}
With this definitions the hamiltonian constraint
\begin{eqnarray}
{\cal H}\mid\psi _F\rangle =[p_T+H(k_a,T)]\psi _F\rangle =0
\end{eqnarray}
becomes a Schroedinger equation
\begin{eqnarray}
-i\frac{\partial}{\partial t}\psi_F (x,t)+
\sum_{x'}(H(t))_{xx'}\psi (x',t)=0
\end{eqnarray}

Such kind of systems will have $f-2$ constants of the motion (or
perennials) $O_i(k_a,t)$ associated to the initial conditions of the
system. They satisfy
\begin{equation}
{i{\partial O}\over{\partial t}} -[H,O]=0    \label{perennial}
\end{equation}

Now, a complete set of compatible perennials define a non degenerate
basis in $E_Q$
\begin{equation}
O_i(t_0)|\alpha_i>=\alpha_i|\alpha_i>
\end{equation}
and
\begin{equation}
|{\psi_F}^{\alpha_i},t>=U(t,t_0)|\alpha_i>|t_0>
\end{equation}
where $U$ is the evolution operator in $E_Q$
\begin{equation}
{i{\partial U}\over{\partial t}} =H U
\end{equation}
The vectors $|{\psi_F}^{\alpha_i},t>$
are eigenvectors of $O_i(t)$ with eigenvalues $\alpha_i$ and span
the physical space.
Thus, there is an isomorphism for any $t_0$
between the restriction ${\cal E}^*_Ft_0$ and
${\cal E}^*_Q$ which is also isomorphic to  ${\cal E}^*_Kt_0$,
spanned by
$|x>|t_0>$.
To conclude in the case of a deparametrizable system, conditions {\bf
(I)} , {\bf (II)} and {\bf(IV)} hold while condition {\bf(III)} is
satisfyed with the usual time independent inner product and the
eigenvectors of the complete set $O_i$ obey the orthonormality
conditions

\begin{equation}
<{\psi_F}_{\alpha_i}|{\psi_F}_{\alpha_j}>_t = <\alpha_i|\alpha_j> =
\delta_{ij} \label{orthono}
\end{equation}
instead of (\ref{ortho}). Thus, all the states are of the same level
$t=t_I$,
taken as the origin of time.

Let us now study the converse. Let $T(q)$ be a time variable
satisfying
conditions {\bf I} to {\bf IV} in the particular case that all the
states are of level $t=t_I$ and the inner product have the form
(\ref{in}).
Let $O_i$ be a complete set of
observables with eigenvectors $|{\psi_F}_{\alpha}>$ that form a
complete
basis on ${\cal E}^*_{Ft} = {\cal E}^*_{Kt}$ and therefore satisfy the
clausure relation
\begin{equation}
\sum_\alpha|{\psi_F}_{\alpha},t>\langle{\psi_F}_{\alpha},t| = I_t
\label{clausure}
\end{equation}
where $I_t$ is the identity operator in $ {\cal E}^*_{Kt}$.
Given an arbitrary state $|\phi\rangle \in {\cal E}^*_{Kt_0}$ , there
is a vector of the physical space $|\phi_F\rangle$
\begin{equation}
|\phi_F\rangle =
\sum_\alpha|{\psi_F}_\alpha\rangle\langle{\psi_F}_\alpha|\phi
\rangle_{t_0}
\end{equation}
such that$|\phi_F,t_0\rangle = |\phi\rangle$.
That implies
\begin{eqnarray}
\phi_F(x,t)&=&\sum_\alpha\sum_{x'}
{\psi_F}_\alpha(x,t){\psi_F}^*_\alpha
(x',t_0)\phi(x',t_0) \nonumber\\
&=& \sum_{x'}D(x,t;x',t_0)\phi(x',t_0) \label{prop}
\end{eqnarray}
Making use of the orthonormality conditions (\ref{ortho}) for vectors of
level $t_I$ one can show that this relation is invertible. Thus one can
write
\begin{eqnarray}
i{{\partial \phi}\over{\partial t}}(x,t)&=&
\sum_{x',x''}i{{\partial D}\over{\partial t}}(x,t;x'',t_0)
D(x'',t_0;x',t)\phi(x',t)\nonumber\\
&=& \sum_{x'}H_{xx'}\phi(x',t) \label{hamilt}
\end{eqnarray}
It is immediate to show from (\ref{prop}) making use of
(\ref{clausure})
and (\ref{ortho}) that the inner product is conserved
$ \langle \psi\mid
\phi\rangle_t=\langle \psi\mid \phi\rangle_{t_0}$, and therefore the
hamiltonian $H$ is hermitic. Thus we recover the Schroedinger
equation for a deparametrizable system.
One can easily prove by making use of (\ref{hamilt}) and the
definition of the perennial operators that they  also satisfy the
evolution equation (\ref{perennial}). Consequently the set of
conditions given in the previous section are a generalization of
the usual quantization procedure for deparametrizable systems. In
the next section we will show several examples of evolving systems.

\section{Applications}
We shall now apply the set of conditions that we have just
established, to
several systems that require an
intrinsic definition of
time. The first and simpler example is the parameterized free
particle in 1+1
dimensions. As a second example we shall consider a discrete
constrained system with an evolving Hilbert space. As a third
example we shall consider a continuous system that behaves as the
continuum limit of the previous model. We shall introduce an
unitary isomorphism that will allow to describe this system in a
fixed Hilbert space. We shall prove that the system still have
perennials with new eigenvalues and eigenvectors at each level $t$
satisfying the generalized orthonormality condition [\ref{ortho}].

\subsection{The parameterized particle}
The parametrized particle obviously is a deparametrizable system
and therefore there is a time variable satisfying conditions
{\bf I} to {\bf IV}. However it is interesting to discuss how these
conditions determine the time variable.
The dynamics of the parameterized particle is contained in the
hamiltonian constraint
\begin{equation}
{\cal H}=P_1+\frac{P_2^2}{2m}=0
\end{equation}
which is quadratic in the momentum $P_2$ .Let us make the natural 
choice $t= x_1$
At the quantum level the kinematical space ${\cal E}_K$ is given in
the basis
$\mid x_1,x_2\rangle$ by functions $\psi (x_1,x_2)$ while the
physical
space is
restricted by the hamiltonian constraint
\begin{equation}
{\cal H}\psi _F(x_1,x_2)=-i\frac{\partial \psi _F}{\partial x_1}-
\frac{1}{2m}
\frac{\partial ^2\psi _F}{\partial x_2^2}=0
\end{equation}
The physical states may be written in term of the Fourier Transform as
\begin{equation}
\psi _F(x_1,x_2)=\frac{1}{\sqrt{2\pi}}\int _{-\infty}^{\infty}dp_2
e^{i(p_2x_2-\omega (p_2)x_1)}f(p_2)
\end{equation}
where
\begin{equation}
\omega (p_2)=\frac{1}{2m}p_2^2
\end{equation}
We introduce the inner product of two physical
states $\psi _1$ and $\psi _2$ in
${\cal E}_{Ft}$
\begin{equation}
\langle\psi _1\mid\psi _2\rangle _t=\int _{t=x_1}dx_2
\psi _1^{\ast}(x_1,x_2)\psi _2(x_1,x_2) .
\end{equation}
 In
this particular
case, the vector spaces ${\cal E}_{Ft}$ for different $t$
are isomorphic. The perennial
operators that commute with the time operator $X_1$ are
\begin{equation}
P_2\,\,\,\, and \,\,\,\, X_2-\frac{P_2}{m}X_1
\end{equation}
and a complete set of commuting perennials is formed by
$P_2$ which obviously is
self adjoint with the inner product (31) and satisfy
condition {\bf II}. Their eigenvectors
\begin{equation}
\psi _{Fp_2}(x_1,x_2)=e^{i(p_2x_2-\omega (p_2)x_1)}
\end{equation}
form an "improper" orthonormal basis of the physical state space,
according with condition {\bf III}.
Finally it is immediate
to check that any square integrable function $\psi (x_1^0,x_2)$
belonging to
${\cal E}_{Kx_1^0}^{\ast}$, the functional space of fixed
$x_1=x_1^0$, may be
expanded in terms of the $\psi _{Fp_2}(x_1^0,x_2)$ that form a
complete basis
of the kinematical space at $x_1^0$. Thus the complete set of
conditions is satisfied
and the usual formalism of quantum mechanics is recovered with $x_1$
as the intrinsic time
of the system.
Before concluding this example, let us briefly mention
what happens if
we take as a "time variable" the coordinate $x_2$. Condition {\bf I}
and {\bf II}  still holds
but one can easily check that {\bf III} and {\bf IV}
cannot be simultaneously satisfied.

\subsection{A relational systems with a finite dimensional
evolving Hilbert space}

In this example we want to show a system where the set of
conditions for
an intrinsic time holds provided its Hilbert space evolves in time.
This kind of model shows how evolving Hilbert spaces arise
within this approach .\\
We consider a system formed by the tensor product of two subsystems
with angular momentum $j$, integer,
\begin{equation}
{\cal E}_K=\{\mid j,m_1\rangle\otimes\mid j,m_2\rangle\equiv
\mid m_1,m_2\rangle
\,\,\, ,\,\,\,\, m_1\geq m_2\}.
\end{equation}
The system is constrained by a hamiltonian constraint
\begin{equation}
{\cal H}=J_0-J_{1}
\end{equation}
where
\begin{eqnarray}
J_0\mid m_1,m_2\rangle=2(j+m_2+1)\mid m_1,m_2+1\rangle\\
J_{1}\mid m_1,m_2\rangle=2(j+m_2)\mid m_1,m_2\rangle
\end{eqnarray}
We introduce a discrete time operator in ${\cal E}_K$ such that
\begin{equation}
T\mid m_1,m_2\rangle=(2j-m_1+m_2)\mid m_1,m_2\rangle
=t\mid m_1,m_2\rangle
\end{equation}
One can easily see that $0\leq t\leq j$. Any state of the basis
$\mid m_1,m_2\rangle$
in ${\cal E}_K$ may be labeled by the value of $t$ and the total
third component
of the angular momentum, $M$.\\
We have
\begin{eqnarray}
\begin{array}{lllll}
t=0 & \mid j,-j\rangle & \Rightarrow & \mid t=0,M=0\rangle &
dim{\cal E}_{K0}=1\\
t=1 &
\begin{array}{l}
\mid j-1,-j\rangle\\
\mid j,-j+1\rangle
\end{array}
&\Rightarrow & \mid t=1,M=\pm 1\rangle & dim{\cal E}_{K1}=2\\
t=2 &
\begin{array}{l}
\mid j-2,-j\rangle\\
\mid j-1,-j+1\rangle\\
\mid j,-j+2\rangle
\end{array}
&\Rightarrow &
\begin{array}{l}
\mid t=2,M=\pm 2\rangle\\
\mid t=2,M=0\rangle
\end{array}
& dim{\cal E}_{K2}=3
\end{array}
\end{eqnarray}
with
$$\hbox{dim} {\cal E}_{Kt}=t+1$$
and $M+t \ge 0$
The physical state space is defined by functions
$\psi _F(t,M)\theta(M+t)$ such that
\begin{equation}
(M+t)[\psi_F(t,M)-\psi_F(t-1,M-1)]=0
\end{equation}
The operator $J_{1z}$, explicitly given by
\begin{equation}
J_{1z}\psi (t,M)=\frac{1}{2}(2j-t+M)\psi (t,M)
\end{equation}
is a perennial and commutes with $T$. The eigenvectors of
$J_{1z}$ define a
basis in the physical state space
\begin{equation}
\psi _{Fm_1}^{j-m_1}(t,M)=\delta _{M,2m_1-2j+t}\theta (t-j+m_1)
\end{equation}
This wave functions vanish for $t<j-m_1$ and therefore they have
level $j-m_1$.\\
The inner product
\begin{equation}
\langle\psi\mid\phi\rangle _t=\sum_{M=-t}^t \psi ^{\ast}(t,M)
\phi (t,M)
\end{equation}
insures the hermiticity of $J_{1z}$, and the inner product between the
elements of the physical basis satisfy the orthonormality condition
\begin{equation}
\langle\psi ^{j-m_1}_{Fm_1}\mid\psi ^{j-m'_1}_{Fm'_1}\rangle _t=
\delta _{m_1m'_1}\theta (t-j+m_1)
\end{equation}
Thus conditions ({\bf I}) to  ({\bf III}) are satisfied, the last
condition also holds,
in fact any function of $M$ at a given $t$ may be obtained by
superposition
of elements of the physical basis $\psi _{Fm_1}^{j-m_1}(t,M)$
at this $t$.\\
Now, it is very easy to compute a transition amplitude
between two eigenstates
of any observable. For instance if we consider the operator
$J_{2z}$ given by
\begin{equation}
J_{2z}\psi (t,M)=\frac{1}{2}(M-2j+t)\psi (t,M)\;,
\end{equation}
it is selfadjoint and commutes with $T$. Now at time $t=1$
$\mid t=1,M=1\rangle$ is an
eigenvector
\begin{equation}
J_{2z}\mid 1,1\rangle =(1-j)\mid 1,1\rangle
\end{equation}
If we label its eigenvectors at time $t$ by
 $\mid m_2,t\rangle$, the transition
amplitude
between this states and $\mid m_2=-j+1,t\rangle$ is given by
\begin{equation}
\langle m_2=-j+1,t=1\mid \mid m_2,t\rangle =\delta _{m_2,-j+t}
\end{equation}

\subsection{An evolving system with an infinite dimensional Hilbert
space}
Evolving systems in continuum space with a continuous
intrinsic time
may be simply introduced. The following example is the continuous
extension of
the model that we have just considered in subsection (3.3)
\begin{equation}
{\cal E}_K=\{\mid a,b\rangle \,\, ,\,\,\, a,b\in R\,\,\,
a\geq b\,\, ,\,\,\,
-1\leq a,b\leq 1\}
\end{equation}
with hamiltonian constraint
\begin{equation}
{\cal H}=4(a-1)(b+1)\frac{\partial}{\partial b}\psi (a,b)=0
\end{equation}
Then if one chooses as a time variable $t=2-a+b$, and $x=a+b$
the hamiltonian constraint takes the form
\begin{equation}
{\cal H}\psi (x,t)=(x^2-t^2)\left[\frac{\partial}{\partial t}
\psi (x,t)+\frac{\partial}{\partial x}\psi (x,t)\right] =0
\end{equation}
A complete set of perennials satisfying condition ({\bf III}) is
given by
\begin{equation}
A\psi (x,t)=a\psi (x,t)
\end{equation}
with eigenvectors belonging to the physical space
given by
\begin{equation}
\psi ^{1-a}_{Fa}(x,t)=\delta (x-t+2-2a)\theta (t-1+a)
\end{equation}
The inner product is given by
\begin{equation}
\langle\psi\mid\phi\rangle _t=\int _{-t}^{+t}dx
\psi ^{\ast}(x,t)\phi (x,t)
\end{equation}
and the eigenfunctions verify the orthogonality condition
\begin{equation}
\langle\psi ^{1-a}_{Fa}\mid\phi ^{1-a'}_{Fa'}\rangle_t=
\frac{1}{2}\delta (a-a')\theta (t-1+a)\theta (t-1+a')
\end{equation}
Thus, contitions ({\bf I}) to ({\bf IV}) obviously hold in this
system that describes  waves propagating in a region bounded
by the future light cone.
By now, we have a description in terms of an inner product with a
time dependent  measure. In the continuous case it is always
possible to introduce an unitary isomorphism between the Hilbert
spaces at two different times. Let us consider this transformation
in the present case.

Let
\begin{equation}
\psi'(x,t) = \sqrt{t}\psi(xt,t)
\end{equation}
then
\begin{eqnarray}
\langle\psi' (t)\mid\phi'(t)\rangle   = \int_{-1}^1 dx
{\psi'}^*(x,t)\phi'(x,t)=\\
\int_{-t}^tdu \psi^*(u,t)\phi(u,t) =
\langle\psi \mid\phi\rangle_t   \nonumber
\end{eqnarray}

where $u=xt$.
Thus the unitary transformation is given by

\begin{equation}
U_t^\dagger(u,x)= \sqrt{t}\delta(u-xt) = U_t(x,u)
\end{equation}
and the eigenvalue equation for the perennial operator  takes the
form

\begin{equation}
A'\psi'_a(x,t)= [{{xt-t}\over2}+1]\psi'_a(x,t) =a\psi'_a(x,t)
\end{equation}

while the the hamiltonian constraint now becomes
\begin{equation}
{\cal H}'\psi'(x,t)=(x^2t^2-t^2)[{(1-x)\over t}{\partial \over
{\partial x}}+
{\partial\over{\partial t}}-\textstyle1/{2t}]\psi'(x,t)=0
\end{equation}

        A complete set of eigenvectors of the perennial operator
belonging to the physical space is

\begin{equation}
{\psi'} ^{1-a}_{Fa}(x,t) =\sqrt{t}\delta[(x-1)t+2-2a]
\end{equation}

Notice that these solutions still have level $1-a$
and verify the orthogonality condition

\begin{equation}
\langle\psi ^{1-a}_{Fa}(t)\mid\phi ^{1-a'}_{Fa'}(t)\rangle=
\frac{1}{2}\delta (a-a')\theta (t-1+a)\theta (t-1+a')
\end{equation}

Thus, we see that the the fundamental properties of the evolving
relational systems the appearance of new eigenvalues and
eigenstates at each level and the conservation of the inner product
among states of level $t_0$ for any $t\ge t_0$, are still present
in this description in terms of a fixed Hilbert space. In general,
evolving Hilbert spaces seem to be naturally related with  systems
with boundaries. For instance the system we have just analized may
be simply generalized to a Klein Gordon system ${\cal
H}={P_t}^2-{P_x}^2$ in a bounded region $R = {(x,t) \; : \; t^2-x^2
\ge 0}$ . This system may be treated as a deparametrizable system
by introducing a time variable $\tau=\sqrt{t^2-x^2}$, leading to a
usual Klein Gordon equation with a nonpositive definite inner
product. However, an equivalent hamiltonian in the bounded region
${\cal H}'= (x+t)({P_t}^2-{P_x}^2)$ may be quantized with a
positive definite inner product in an evolving Hilbert space.

\section{Conclusions and Final Remarks}

We have introduced a notion of intrinsic time in relational systems
that allow to recover the fundamental features of time in quantum
mechanics. In the case of any standard quantum mechanical systems
in parameterized form our method reproduce the usual formalism of
quantum mechanics.\\

However the method allows to include relational dynamical systems
and leads naturally to quantum mechanical systems with an evolving
Hilbert space. In that sense we are implementing the intuition that
one can define fixed time Hilbert spaces that contain subsets of
all possible states of the system. These systems are not invariant
under time translations or time reversal and have a defined arrow
of time. The initial state of the system, as well as the evolution
of the Hilbert space, are determined by the hamiltonian constraint
and therefore dictated by the dynamics. The number of accessible
states increases in time.\\

Let us conclude with some final comments about the quantum gravity
case. Even though very little is known about the physical state
space of quantum gravity, a pure gravity system could behave as an
evolving system of this type. In fact, it is natural to take as the
configuration space of quantum general relativity the loop space
\cite{RovSmo1}, because in this representation the domain of the
wave functions seems to be simply related with the microscopic
structure of space time. In the loop representation, the
kinematical space ${\cal E}_K$ is given by the knot dependent
functions \cite{RovSmo1} $\psi [K]$ that satisfy the diffeomorphism
constraint. As a candidate for the intrinsic time $t$, we would
like to take a variable such that the simplest configuration
correspond to its initial value. A good candidate seem to be the
minimum number of crossings of a knot, this knot invariant quantity
may be used to characterize the complexity of each knot. The
kinematical space of quantum gravity will be characterized by wave
functions $\psi(t,\kappa)$ where $\kappa$ are the remaining knot
invariants necessary to describe a knot with minimum number of
crossings equal to $t$. If we do not include knots with more than
triple selfintersections, the number of independent knot invariants
with a fixed $t$ is finite and increases with $t$ Of course, if we
want to take as a time variable some knot invariant such that the
number of independent knots and the dimension of the kinematical
space increases with $t$, there is not a unique choice. For
instance, one could define as time the degree of an universal
polynomial associated with the link. However, up to now, it is not
known how to classify any knot in terms of knot polynomials in such
a way that  inequivalent knots always correspond to different
polynomials.

In general relativity no perennial is known, but a candidate to
observable, in the  sense used in this paper, is given by the
volume of the Universe. The eigenstates of the volume operator are
knot states having a definite number of crossings and intersections
and its eigenvalues  are essentially proportional to the Planck
volume times the number of intersections. This operator does
commute with the diffeomorphism constraint and with our "time" and
does not commute with the hamiltonian constraint. These are  the
conditions  required to our observables. The naive picture of the
Big Bang that we get is a unique zero volume state that evolves
with certain probabilities to different states of finite volume.
Within this description the recollapse of the Universe will be
associated with a decreasing  volume while the complexity of the
knot space is still growing. Unfortunately, it does not seem to be
easy to check a proposal of this type on a simple cosmological
model. In fact, in that case the knot structure related with the
diffeomorphism invariance is not present.

{\bf Acknowledgements}\\

We wish to thank to Daniel Armand-Ugon, Alcides Garat, Don Marolf,
Jose Mourao  Jorge Pullin and Lee Smolin for  fruitful discussions.
One of us R.G. wish to thank the grant NSF-INT-9406269
for partial support.


{\small
\begin{thebibliography}{99}
\bibitem{von} J. von Neumann, {\it Mathematical Foundations of
Quantum Mechanichs} (Princeton University Press, Princeton,
N.J.,1955)
\bibitem{Dirac} P.A.M. Dirac, Lectures on quantum mechanics,
{\it Belfer 
Graduate
School of Science Monographs, n$^{\underline{o}}$2},
(Yeshiva University Press, New York, 1964)
\bibitem{sunder} K. Sundermeyer {\it Constrained Dynamics},
(Lecture Notes in Physics 169 Springer-Verlag Berlin 1982)
\bibitem{Kuc1} K. Kuchar, Time and interpretations of quantum
gravity,
{\it Proceedings of the 4$^{th}$ Canadian Conference on General
Relativity
and Astrophysics}, ed. G. Kunstatter {\it et al.}, (World Scientific,
Singapore, 1992)

\bibitem{Isham1} C. Isham, Canonical quantum gravity and the problem
of time,
{\it Imperial College Preprint} Imperial/TP/91-92/25 (1992)

\bibitem{Kuc2} K. Kuchar, Canonical quantum gravity,
{\it Proceedings of the 
13$^{th}$
International Conference on General Relativity and Gravitation},
ed. R. 
Gleiser {et al.},
(Institute of Physics Publishing, Bristol and Philadelphia, 1992)

\bibitem{UnWa} W. Unruh and R. Wald, Phys. Rev. D 40 (1989) 2598


\bibitem{Ash} A. Ashtekar, Phys. Rev. Lett. 57 (1986) 2244

\bibitem{Unruh} W. Unruh, Time in quantum gravity and quantum
mechanics, {\it 
British Columbia
University Preprint} (1993)

\bibitem{Jac} T. Jacobson, Black hole thermodynamics and the space
time 
discontinuum,
\\{\it Conceptual Problems of Quantum Gravity}, ed. A. Ashtekar
{\it et al.},
(Birkh$\ddot{a}$user,
Boston, 1991)

\bibitem{Hartle} J. Hartle, Time and predictions in quantum
cosmology,
\\{\it Conceptual Problems of Quantum Gravity}, ed. A. Ashtekar
{\it et al.},
(Birkh$\ddot{a}$user,
Boston, 1991)

\bibitem{RovSmo1} C. Rovelli and L. Smolin, Nucl. Phys. B 133
(1990) 80


\bibitem{Smo} L. Smolin, Space and time in the quantum universe,
\\{\it Conceptual Problems of Quantum Gravity}, ed. A. Ashtekar
{\it et al.},
(Birkh$\ddot{a}$user,
Boston, 1991)




\end{thebibliography}}
\end{document}